# Multi-scale super-resolution generation of low-resolution scanned pathological images

Kai Sun[1], Yanhua Gao[2], Ting Xie[1], Xun Wang[1], Qingqing Yang[1], Le Chen[1],

Kuansong Wang[3], Gang Yu[1]

1. Department of Biomedical Engineering, School of Basic Medical Sciences, Central South University, #172 Tongzipo Road, Changsha, 410013, China.
2. Department of Ultrasound, Shaanxi Provincial People's Hospital, #256 Youyixi Road, Xi'an, 710068, China.
3. Department of Pathology, School of Basic Medical Sciences, Central South University, #172 Tongzipo Road, Changsha, 410013, China.

Co-corresponding author：
Gang Yu, Ph.D.
School of Basic Medical Science
Central South University
Changsha, P.R. China
Email: yugang@mail.csu.edu.cn

**Background.** Digital pathology has aroused widespread interest in modern pathology. The key of digitalization is to scan the whole slide image (WSI) at high magnification. The lager the magnification is, the richer details WSI will provide, but the scanning time is longer and the file size of obtained is larger. For example, the file size of each WSI of 40 times magnification (40X) exceeds 1 Gigabyte, which leads to huge storage capacity, very slow scanning and network exchange, seriously increasing time and storage costs for digital pathology.

**Methods**. We design a strategy to scan slides with low resolution (5X) and a super-resolution method is proposed to restore the image details when in diagnosis. The method is based on a multi-scale generative adversarial network, which sequentially generates three high-resolution images such as 10X, 20X and 40X. A dataset consisting of 100,000 pathological images from 10 types of human body systems is performed for training and testing. The differences between the generated images and the real images have been extensively evaluated using quantitative evaluation, visual inspection, medical scoring and diagnosis.

**Results.** The file size of each 5X WSI is about 15 Megabyte. The peak-signal-to-noise-ratio of 10X to 40X generated images are 24.16, 22.27 and 20.44, and the structural-similarity-index are 0.845, 0.680 and 0.512, which are better than other super-resolution networks. Visual inspections show the generated images have details similar to the real images. Visual scoring average and standard deviation from three pathologists is 3.63±0.52, 3.70±0.57 and 3.74±0.56 and the p value of analysis of variance is 0.37, indicating the pathologists confirm that generated images include sufficient information for diagnosis. The average value of Kappa test of the diagnoses of paired generated and real images from 49 diseases of 10 body systems is 0.99, meaning the diagnosis of generated images is highly consistent with that of the real images.

**Conclusion**. This proposed method can generate high-quality 10X, 20X, 40X images from 5X images at the same time, in which the time and storage costs of digitalization can be effectively reduced up to 1/64 of the previous costs. These generated high-resolution images can replace the real images for clinical diagnosis, which provides a better alternative for low-cost storage, faster image share of digital pathology.

# 1 Introduction

In the pathological examination, technicians make glass slides of human tissue, such as routine haematoxylin and eosin (HE)[1], frozen section[2][3] and special staining[4]. The pathologist magnifies the slides with a microscope to diagnose the disease. In recent years, the digitalization of pathological slides has begun to be promoted because of the convenient storage, network transmission and effective computer aided diagnosis [5][6][7].

A high-resolution (HR) scanner with 20X or even 40X lens is used to scan the glass slide and the WSI is obtained [8]. Because of the magnification of dozens of times, the pixels in one WSI are up to 50000*50000 pixels [9], and the file size is more than 1 Gigabyte [10], which greatly increases the storge costs and takes a few minutes for a network exchange [11]. Taking China as an example, there are up 50 million new pathological examinations every year, which requires at 50 million Gigabyte storage capacity. Moreover, the HR scanning of each WSI needs a few minutes or more [12], so the digitization of the WSIs requires more than 200 years of scanning time. In addition, because the depth of field of the HR lens is very small, the tissue on the slide must be very flat or parts of the WSI may be blurred [13]. Therefore, the digitalization of pathological glass slides is inefficient and costly.

We assume that the low-resolution (LR) scanning such as 5X magnification can provide an alternative strategy. Compared with 40X scanning, the speed of 5X scanning increased by 64 times, and the size of WSI reduces to 1/64. Secondly, the depth of field of LR lens is larger, which could reduce requirements of the flatness of human tissues. However, the 5X images have lost enough details for diagnosis, so a method must be proposed to restore the missing details and generate a series of high-resolution images from 5X images.

The conventional methods of obtaining HR images from LR images are the algorithms such as bilinear interpolation and B-spline interpolation [14]. However, it is difficult for them to recover the lost details. Recently, in order to get high-quality images from LR images, the super-resolution (SR) methods based on deep learning have been widely studied in natural images. There are many excellent super-resolution algorithms, for example, deep recursive residual network (DRRN) [15], super-resolution generative adversarial network (SRGAN) [16], progressive generative adversarial networks (P-GANs) [17], and multi-scale information cross-fusion network (MSICF) [18]. The SR methods are also used in medical images such as magnetic resonance [19] [20], computer tomography [21], and ultrasound images [22]. For example, the super-resolution convolutional neural network (SRCNN) is used for the computer tomography images [23], and the pathological super-resolution model for cytopathological images [24].

Although the SR methods have been successfully applied in many fields, the pathological images are different from natural and other medical images. There are a large number of cells and rich micro-tissue structures in them, but general SR methods pay attention to the larger image texture and structure. Moreover, the maximum magnification of almost all SR methods is 4 times [25], but the magnification of the pathological image is up to 8 times, e.g., from 5X to 40X. For the convenience of diagnosis, a series of HR images should be continuously obtained from 5X images, however, the magnification of almost all SR methods is fixed. We propose a multi-scale SR method (MSR), which continuously generates 10X, 20X, and 40X HR images from 5X images. The highlights of the proposed method are as follows:

(1) The size of the 5X scanned image is small enough, about 15 Megabyte, and only one-64th of the image file the scanned by 40X. The presented method can generate 10X, 20X and 40X HR images at the same time. The optimization is applied on three resolutions, and the generated images are of realistic and rich details, better than the existing methods.

(2) The differences of generated images and real images are evaluated on pathological diagnosis of 10 types of human body systems, which has proved that generated images can replace real HR images for pathological diagnosis. Therefore, MSR can provide a low-cost method for pathological digitalization.

## 2 Methodology

**Dataset**

The pathological glass slides are from Xiang-ya Hospital of Central South University, which is one of the most famous hospitals in China. The technicians randomly selected 200 whole slides, and the slides are composed of 10 types of human body systems, each of which has 20 slides. The slides were scanned by using KF-PRO-005 digital pathology scanner (KFBIO company, Ningbo City, China) at 40X magnification, and their WSIs were obtained. Refer to Figure 1 for the number of WSIs of 10 human body systems and Appendix A for the pathological diagnosis. Two pathologists reviewed the image quality of WSIs to ensure that they are suitable for pathological diagnosis.

The WSIs were randomly divided into the training, testing and extended testing sets, as detailed in Figure 1. The training set consists of 80 WSIs, which come from 10 types of human body systems, and each system has 8 WSIs. The testing and medical testing sets consist of 20 and 100 WSIs respectively. Because the size of one WSI is too large to be inputted into subsequent deep neural network, the 1,000 non-overlapping images (tiles or patches) in every WSI in training and testing set are randomly taken out, and the size of every image is 1024*1024 pixels(40X). There are a total of 100,000 images in training and testing sets. Two regions of interesting (ROI)/images related to diagnosis from every WSI in medical testing set are selected by a pathologist, and used for medical evaluation.

In order to get LR images from the 40X images, the bicubic interpolation algorithm is used, where the continuous 2 times down-sample from each 40X image to obtain its LR images of 20X, 10X and 5X [26][27], which are 512*512, 256*256 and 128*128 pixels respectively. Therefore, the 5X-20X images in the training and testing sets are obtained, which are treated as the real images.

**Network Architecture**

The proposed network includes three super-resolution blocks (SRBs), which enlarges the input image twice. The 5X LR image is inputted into the network, and the 10X, 20X and 40X SR images are generated by the three SRBs continuously, as shown in Figure 2. The generated images are compared with real images on three resolutions, and the pixel differences (generator loss), feature differences (perceptual loss) and image differences (discriminator loss) are weighted together to guide the training of the network.

The structures of these SRBs are the same, as shown in Figure 3, where the input LR image is firstly convolved by 64 kernels with 3*3 size, and the 10 repeated basic blocks are concatenated to predict the details in SR image. The predicted details are mixed by a convolutional layer, and then added to the input LR image. The sum of predicted and original details is up-sampled by the pixel shuffling operation, and mixed by the final convolution layer to generate the SR image with 3 color channels and 2 times magnification. Meanwhile，every basic block includes three dense blocks, where five dense convolutional layers are concatenated to find multi-scale features from the LR image to reconstruct the lost details.

The loss functions including generator, perceptual and discriminator loss are based on three levels such as pixel, feature and whole image. The generator loss compares the differences of the pixels one by one of the generated images and real images, and is defined as the mean absolute error.

$$GL_{X_j} = \frac{1}{m}\sum_{i=1}^{m}\left|SR_{X_j,i} - HR_{X_j,i}\right| \quad X_j = 10X, 20X, 40X \quad (1)$$

where $X_j$ is the magnification, and $m$ is the pixel number in the image. The SR is the generated image, and HR is the real image.

The high-level representations of generated SR images and real HR images are extracted, where a network of 19 layers from the Visual Geometry Group (VGG19) [28] is selected as the extractor because of its excellent performance on representation learning. The middle-level (layer 5) and high-level (layer 9) representations are compared here for the perceptual loss of the generated image and real image, defined as the equation (2).

$$PL_{X_j} = MSE\left[\phi_5\left(SR_{X_j}\right) - \phi_5\left(HR_{X_j}\right)\right] + MSE\left[\phi_9\left(SR_{X_j}\right) - \phi_9\left(HR_{X_j}\right)\right] \quad (2)$$

where $X_j = 10X, 20X, 40X$, the MSE is the mean square error function, and the $\phi$ is the extractor of VGG19.

In order to better make the generated images have a high degree of authenticity, the discriminator network is used to identify the True or False of the real image and the generated image. Because the discriminator network will be trained with the generator, we only use one discriminator to compare the authenticity of the images with the maximum 40X magnification to reduce the training cost. Generally, the greater the magnification, the more likely the generated image is to be distorted. So, if the quality of the 40X image is good, 20X and 10X should be good too. The discriminator loss is defined as the equation (3).

$$DL_{40X} = -E[\log(D(HR_{40X}))] - E[\log(1 - D(SR_{40X}))] \quad (3)$$

Because there are so many details in 40X pathological images, a very deep discriminator network is used where there are 7 convolutional modules and two fully connected layers. Every convolutional module includes some convolution layers with the 64-512 convolutional kernels with 3x3 size, and the stride size is 1-2. The sigmoid function is connected to the end, outputting the True (1) and False (0), shown in Figure 4.

The two or three losses on each magnification are added, and then weighted together for the total loss function.

$$Loss = \sum_{X_j} w_{X_j} Loss_{X_j} \quad (4)$$

where the $Loss_{X_j}$ is the sum of generator, perceptual and discriminator loss on the magnification $X_j$, $w_{X_j}$ is the weighted value.

**Network Training**

The proposed network is implemented using TensorFlow framework (the version is 1.15.0) [29] and trained on a server consisting of two graphics processing units (GPU) of Tesla V100 32GB, NVIDIA company. The pretrained model of VGG19 is download from the model zoo of TensorFlow and keeps fixed while training. The generator and discriminator are initialized randomly and trained alternately. 10% of the images in the training set are used as the validation set for hyperparameter selection.

The Adam optimizer is selected for the optimization of network weights. The initial learning rate is 0.0004, and decays exponentially where the rate is reduced by 0.5 times every 30 epochs. The total training epochs is 120, and 500 steps is performed every epoch. The batch size is fixed to 2 considering the GPU memory and training costs. The 50 smaller patches with the size of $64 \times 64$ pixels are randomly selected from the input LR $128 \times 128$ image, and the small patches whose standard deviations are bigger than the flatness value are kept, and then input to the proposed network for training. The flatness is the parameter controlling the complexity of the input patches, and only the patches with complex details should be used as the training data. The flatness increases 0.01 from 0 every 5 epochs, but not more than 0.15.

The weights of the three losses are determined after hyperparameter selection, where the weight of the generator loss is 0.06, perceptual loss 0.083, and discriminator loss 0.04. The hyperparameters are listed in Table 1. The code of our method can be downloaded from the link: http://github.com/CSU-BME/pathology_MSR.

## 3 Results

**Peak-signal-to-noise-ratios and structural similarity index**

The quantitative experiments are first described here, and the 10X, 20X and 40X SR images are generated from 5X images in the testing set by the proposed MSR method. Two quantitative indices are computed to evaluate the generation quality. The first one is peak-signal-to-noise-ratios (PSNR), which is defined as the signal difference between generated images and real images of the same magnification [30]. The second is the structural similarity index (SSIM), which is defined as the difference in human vision between generated images and real images [31]. Generally, the larger these two indicators are, the closer the generated image is to the real image.

The 10X-40X SR images are generated from 20,000 5X images in the testing set and compared with real 10-40X HR images. The mean PSNR and SSIM between the generated and real images at different resolutions are shown in Figure 5. The mean PSNR of 10X, 20X, and 40X generated images are 24.167, 22.272, and 20.436 respectively, while the mean SSIM are 0.845, 0.680 and 0.512.

In order to demonstrate our model, the recent other SR models such as DBPN (Deep Back-Projection Networks) [32], ESPCN (Efficient Sub-Pixel Convolutional Neural Network) [33], EDSR (Enhanced Deep Super-Resolution Network) [34], MDSR (Multi-scale Deep Super-Resolution System) [34], RDN (Residual Dense Network) [34] are re-trained and re-tested in our training and testing sets. Because these models support only a fixed magnification, three versions with 2 times, 4 times, and 8 times magnification of each model are trained and tested respectively, for example, DBPN-10X, DBPN-20X and DBPN-40X, which respectively realize the generation of 10X, 20X and 40X images from 5X images in the testing set. Similarly, after the PSNR and SSIM of the 10X, 20X and 40X images generated by each model and the real 10X, 20X and 40X images in the testing set are computed, all PSNRs and SSIMs are averaged at different resolutions. The codes of these models are downloaded from the links given in their published papers, and then retrained with 2 times, 4 times and 8 times magnification according to the provided hyperparameters in the codes. The RDN code produced some errors when magnifying the 5X image by 8 times [34], and the author of the RDN model didn't give a solution, so the PSNR and SSIM of RDN at 40X images cannot be provided.

From Figure 5, it can be seen that the SSIM and PSNR of the 10X images of all methods are the highest, which means that the images generated by 2 times magnification are the closest to the real images, and the image quality is the highest. Because 40X images are down-sampled to 5X images, and most details are lost, so the quality of the generated 40X images is relatively poor. In contrast, our method has higher mean PSNR and SSIM than the other 5 models at 10X, 20X and 40X generated images, which shows that our method has a better restoration of image details. Although the mean SSIM of our method at 40X generated images don't achieve the highest value, it is particularly worth noting that these five models are single-scale, using three versions to achieve 10X, 20X, and 40X image generation respectively.

**The comparison of visual inspections**

Although the mean PSNR and SSIM can evaluate the difference in signal and vision between the generated images and the real images, they are still far away from human visual perception. In Figure 6, a real HR 10X is displayed, in which the region of interest (ROI) is drawn by a green bounding box, and shown in the rightest. The real 5X image is obtained by the 2 times down-sample version of the ROI. The generated 10X images from real 5X images by our model and the other 5 models are shown on the right side. By visually inspecting the generated images of these 6 models, it can be found that all the generated images are of high quality and the image details are quite realistic.

The 20X generated images and real images are also shown. It can be found that our model is relatively close to the real image, but other models have different degrees of distortion. Moreover, the generated 40X images by other models have appeared blurred, while our model can preserve important image structures. It's worth noting that from 5X to 40X, some backgrounds are too small to be completely restored, which is consistent with the PSNR and SSIM.

Local visual comparisons of the generated images by our model and real images have been shown in Figure 7, where our method recovers HR images close to real images. It is worth noting that the background color of the generated image deviates a little from the real image.

**The comparison of pathological diagnosis**

Another 100 WSIs are used to the evaluation of pathological diagnosis by pathologists, as shown in Figure 8. A pathologist with 15 years experiences carefully selects two non-overlapping ROIs (HR ROI-1, ROI-2) closely related to diagnosis from every WSI, and total 100 paired ROIs are obtained. The HR ROI-1 is down-sampled to 5X LR, and then inputted into the presented method (MSR). The SR ROI-1 is generated.

The first comparison is visual scoring of HR ROI-1 and SR ROI-1. Three pathologists (A-C) carefully observe the two paired images, and then give a score on the diagnostic consistency of the two images, as shown in Table 2. Each pathologist scores 100 pairs of HR ROI-1 and SR ROI-1 and the specific scoring situation of each pathologist is shown in Appendix B. The p value of analysis of variance is 0.37, meaning the scores of the three pathologists are not significantly different. The average scores are shown in Figure 9. The visual scoring average±standard deviation from pathologist A, B and C is 3.63±0.52, 3.70±0.57 and 3.74±0.56, indicating that there is no difference in diagnosis between the generated image and the real image.

The 100 paired SR ROI-1 and HR ROI-2 are randomly shuffled, and included in the images of routine diagnosis. Except for the human body system from which the images come, other information, including the WSIs to which the ROIs belong, has been eliminated. Another three pathologists (D-F) blindly review the SR ROI-1 and HR ROI-2 individually. The diagnosis of two ROIs from the same WSI is highly consistent, as shown in Figure 10. The Kappa test shows that $K_D$ about the 100 paired SR ROI-1 and HR ROI-2 from pathologist D is 1.00, $K_E$ from pathologist E is 0.99, $K_F$ from pathologist F is 0.98. The Kappa test demonstrates that there is no statistical difference in diagnosis of three pathologists between the generated images and the real images.

## 4 Discussion

Pathological slides are not conducive to preservation and inconvenient to search and exchange. With the advent of high-resolution scanners, the pathological images from glass slides have applied in modern pathology. The glass slides are scanned into digital WSIs, which can be conveniently stored and transmitted over the network for remote diagnosis. Because there are a lot of microscopic cells and tissues in pathological slides, the HR scanning is necessary in order to retain these details in images. However，the file size of HR scanning is too large, which severely affects the acceptance of digital pathology because of the slow scanning and data exchange, and huge storage costs.

We design an alternative strategy for LR scanning, where the 5X images are only about 15MB, which can completely solve the storage and exchange problems of HR images. Moreover, The LR scanning will reduce the flatness requirement of the human tissue in the slides. However, an accurate diagnosis cannot be made on 5X images, so a method is proposed to convert 5X images to HR images, such as 10X, 20X, and 40X while diagnosis.

Although the existing SR models have been widely used in natural images and medical images, the main problem is the fixed magnification, and the magnification is not more than 4 times magnification. According to the needs of pathological diagnosis, the SR method must be magnified continuously, and the maximum magnification can reach 8 times. The proposed method meets the above requirements and can continuously generate 10X, 20X, and 40X images.

We compare our method with several recent models. Because the other models have only fixed magnifications, we trained three versions respectively, i.e. the generations of 5X-10X, 5X-20X, 5X-40X. In theory, a model that considers a single magnification ratio should perform better than a model that generates multiple magnification ratios at the same time because the learning task is easier. However, our method has achieved better results than those methods in both quantitative evaluation and visual comparison experiments. The possible reason is that the generated image by our method is optimized with the real image in multiple resolution/magnifications, but other methods only consider the optimization at the single magnification.

The visual comparison shows that our method has achieved good results that are significantly better than those of the existing methods, especially the restoration of cell structure and clusters related to the pathological diagnosis. However, it is worth noting that PSNR and SSIM of all methods are lower than those of the natural images. Although our method can restore local details very well, but the color of the generated image is slightly changed. This may be because the method pays attention to the micro-structures such as cells in the pathological image, but not the macro-structure such as the color and background of the image. The PSNR and SSIM mainly consider the differences in macro structures, so the two indicators are not too high.

For application in pathological domains, whether the generated image can replace the real image, the most important is whether it changes the diagnosis result. Therefore, medical confirmation is more important. In the visual assessment, the pathologist believes that although there are visual differences between the generated image and the real image, the information required for diagnosis is complete. Moreover, the blind diagnosis of pathologists on 49 diseases of 10 body systems further confirms the conclusion that there is no statistical difference between the diagnosis of generated images and real images.

The work in this paper proves after the use of low-resolution scanning, the multi-scale method can generate the SR images that retain sufficient information for pathological diagnosis, which can be used as a powerful alternative to solve the problems of high-cost digitalization such as slow scanning of slides and slow sharing of data. In future work, the quality of the SR images should be further improved to have a higher similarity with the real images.

## 5 Conclusion

Digital pathology can easily store and exchange WSIs, but the HR scan will produce large image files, high storage costs and slow network transmission. This paper proposes to scan pathological images with low resolution and use a multi-scale method to generate 10X, 20X and 40X HR images from 5X images. Quantitative and visual experiments have shown that the proposed method can achieve good image quality, and medical experiments further confirm that there is no statistical difference between the diagnosis of generated images and real images. The presented method can be used as a powerful alternative, which is helpful to reduce the cost of digital pathology.

**Acknowledgement**

G.Y. was supported by the Emergency Management Science and Technology Project of Hunan Province (#2020YJ004, #2021-QYC-10050-26366).

Table 1. Hyperparameters used in our model

| Hyperparameters | Value |
|---|---|
| **Epochs** | 120 |
| **Steps per epoch** | 500 |
| **Batch size** | 2 |
| **Optimizer** | Adam |
| **Learning rate** | 0.0004 |
| **Decay factor** | 0.5 |
| **Decay frequency** | 30 |
| **Min flatness** | 0 |
| **Max flatness** | 0.15 |
| **Increase factor** | 0.01 |
| **Increase frequency** | 5 |
| **Generator loss weight** | 0.06 |
| **Perceptual loss weight** | 0.083 |
| **Discriminator loss weight** | 0.04 |

Table 2. Visual scoring rules of generated images by pathologists

| Score | Descriptions |
|---|---|
| 5 | The differences between the generated and real images are so small that they are almost invisible. |
| 4 | There is a slight difference, but it has no effect on the diagnosis. |
| 3 | There is a certain difference, but it has no effect on the diagnosis. |
| 2 | There are obvious differences, which may affect the diagnosis. |
| 1 | The difference is large enough not to be used for diagnosis. |

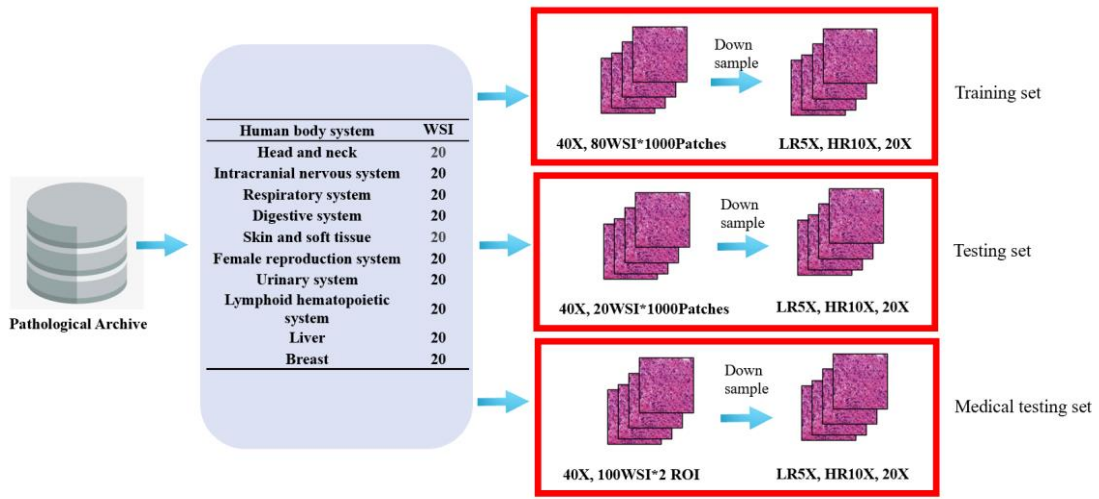

Figure 1. Datasets. The 40X WSIs of 200 (20 of each human body system) are randomly selected from pathological archive. The 100 WSIs are used for training and testing set. Another 100 WSIs are used for medical testing set.

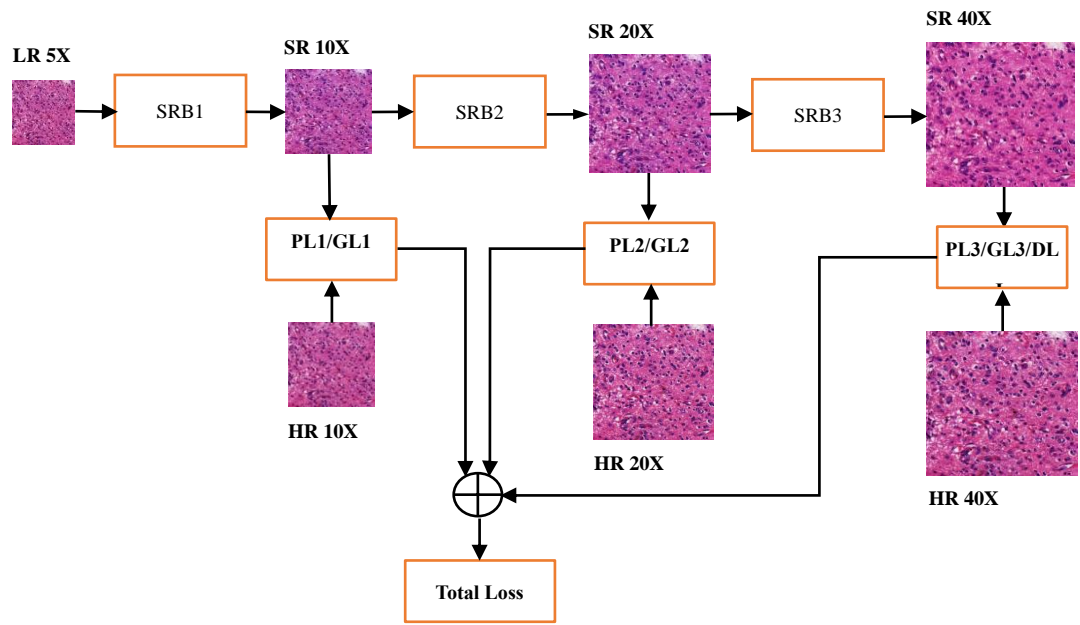

Figure 2. The flowchart of network pipeline. The LR 5X images in training set are inputted the network, where the three SRBs are connected together, and each one generates one SR image. The three generated images are compared with real HR images, and the loss is calculated for training the network, where PL is perceptual loss, GL is generator loss, and DL is discriminator loss.

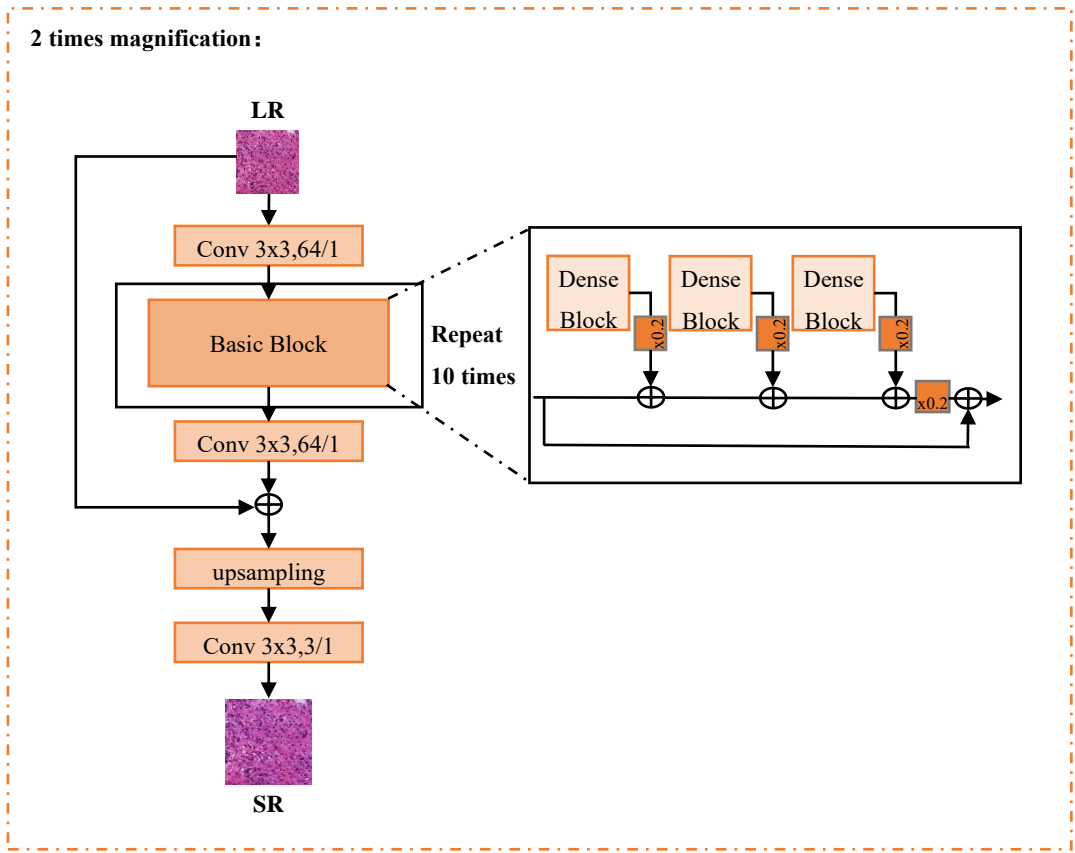

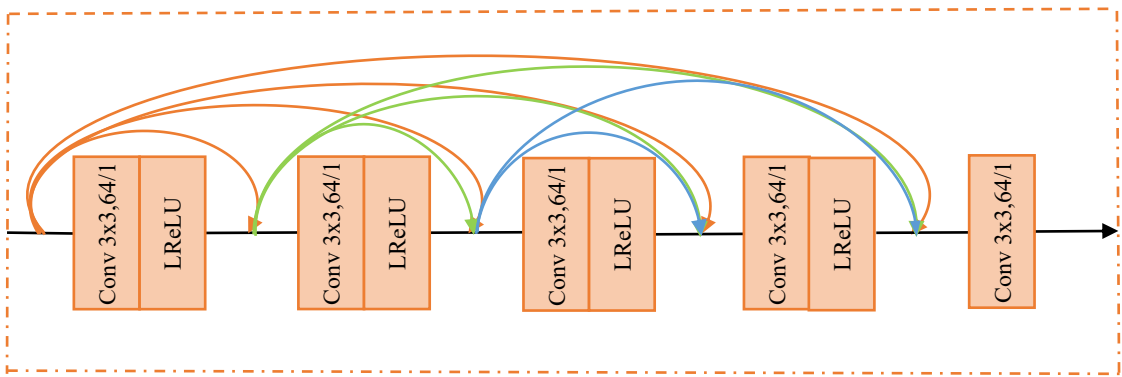

Figure 3. The structure of the SRB, where the convolutional layer is denoted as conv, kernel size, and the number of channels/stride size, the LReLU is the activation function, the upsampling layer is used for pixel shuffling to enlarge the image. The 0.2 is the residual scaling parameter.

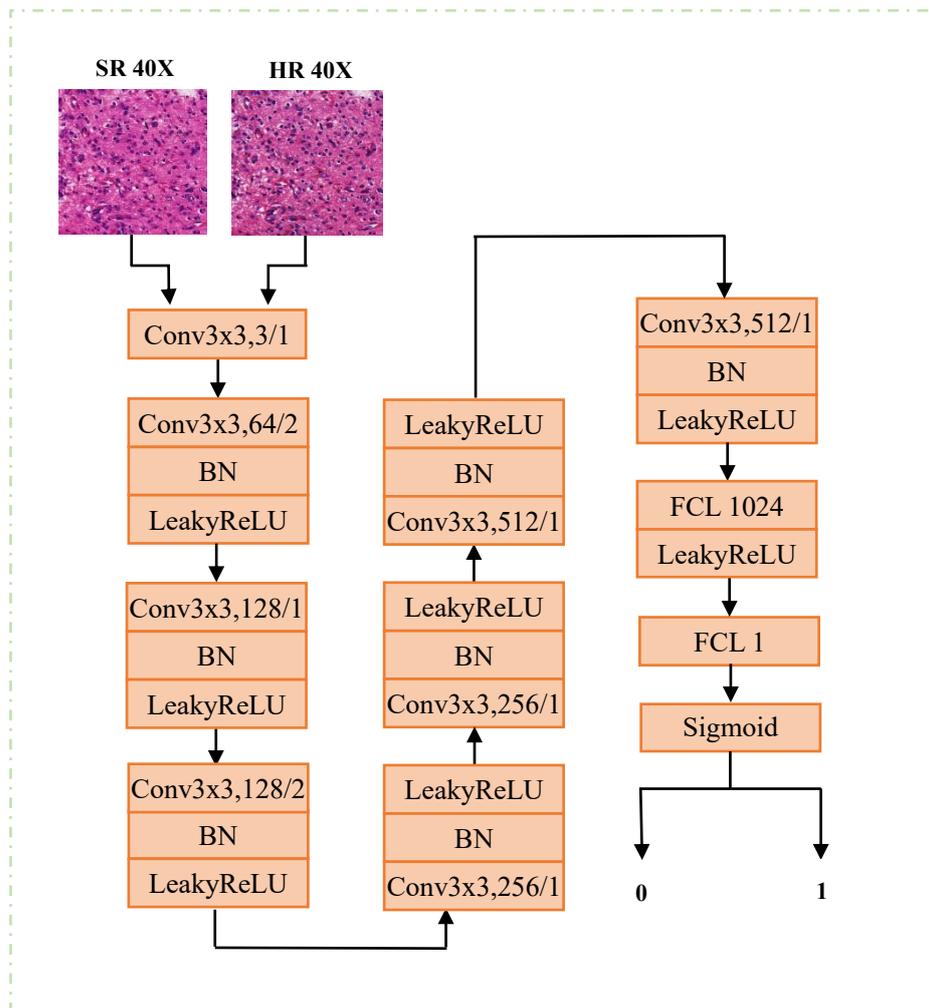

Figure 4. The network framework of the discriminator, where the BN is the batch normalization layer, The FC n is the fully connected layer with output node of n.

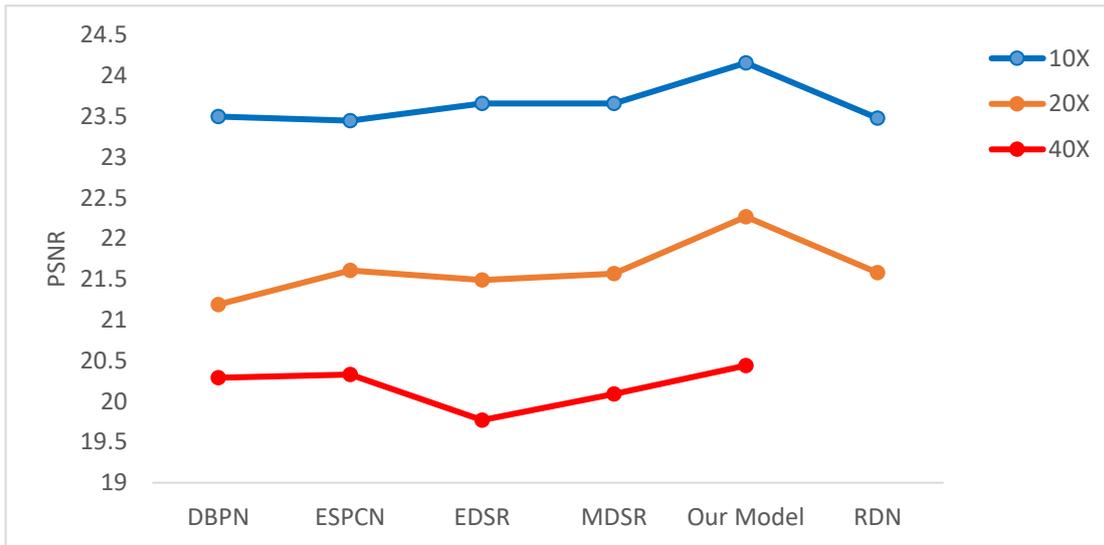

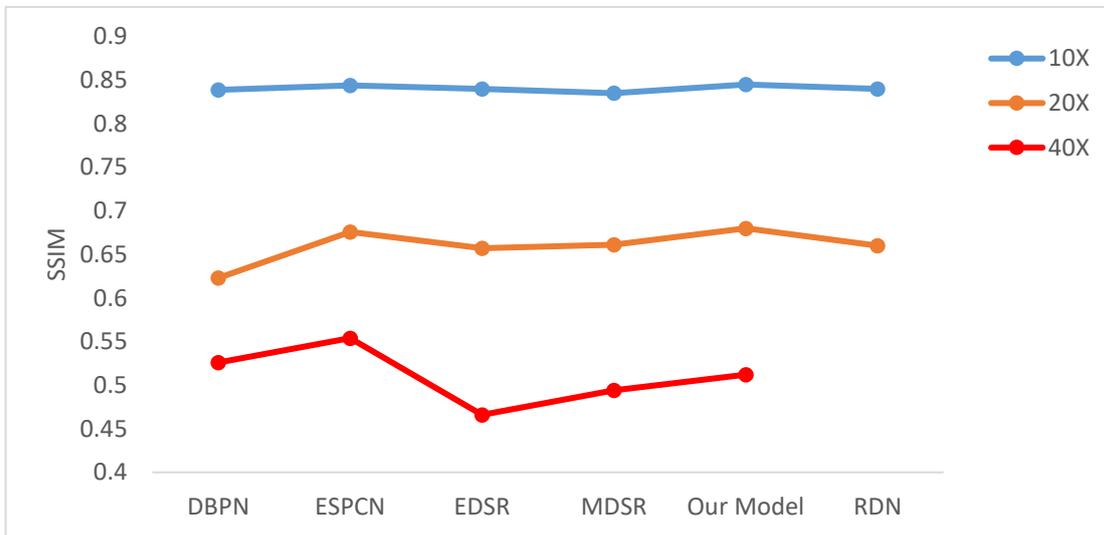

Figure 5. The mean PSNR and SSIM of our model and five recent models at different resolutions in testing set, where three color lines demonstrate

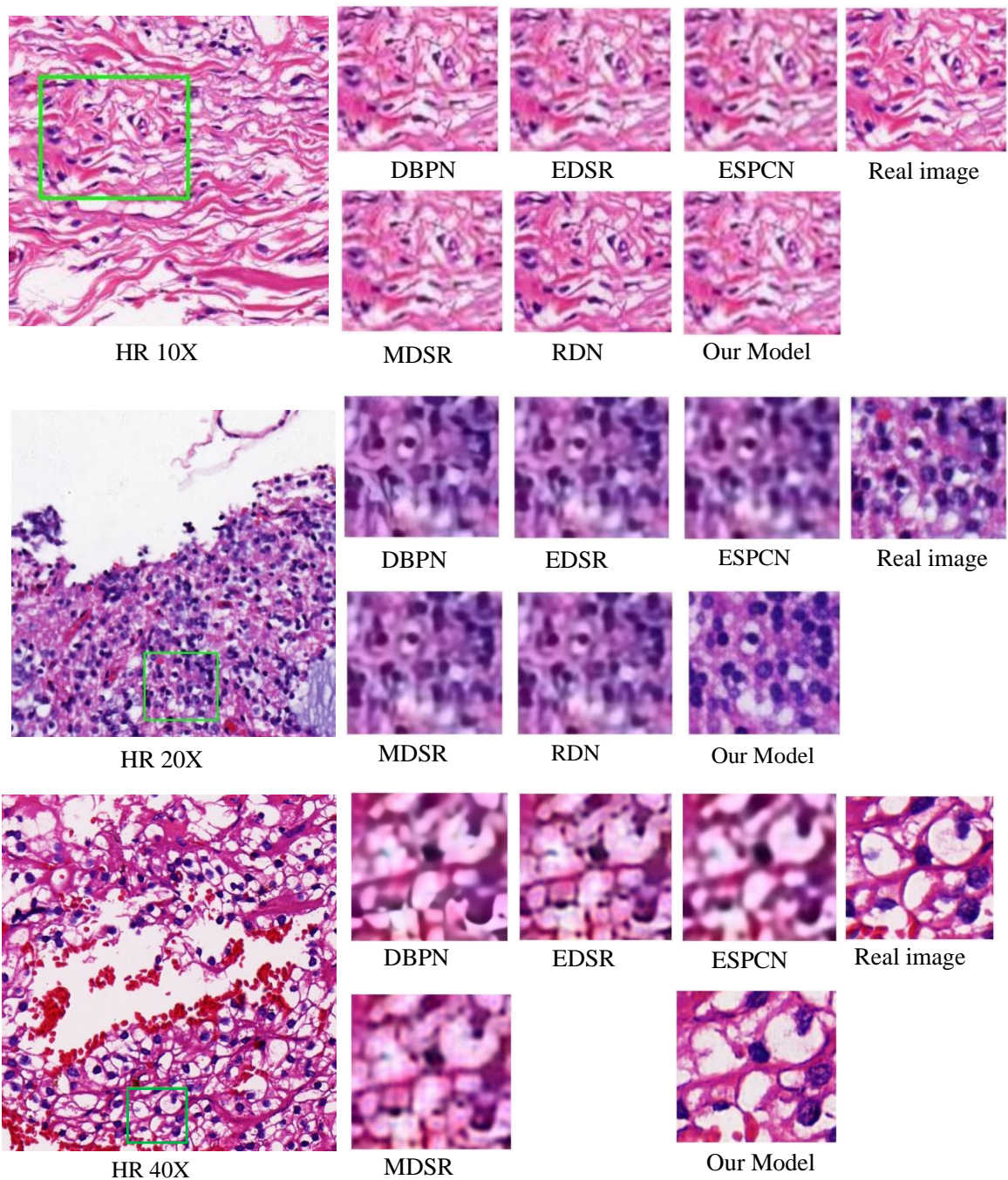

Figure 6. From top to bottom, visual inspections of 10X-40X generated images from 5X image and 10X-40X real image. The left side is real 10X-40X images, while the right side is the generated 10X-40X images.

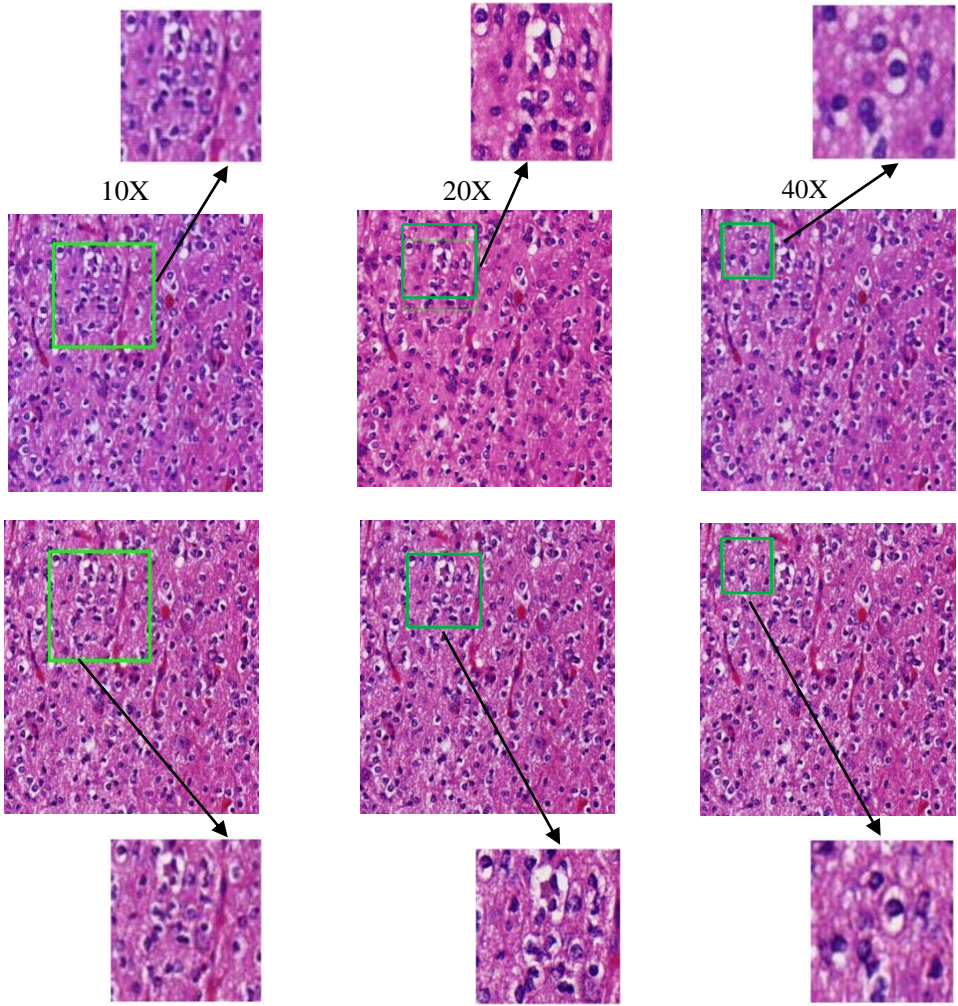

Figure 7. The local visual comparisons of the generated images by our model and real images. The first row is the generated images, and the second row is the real images.

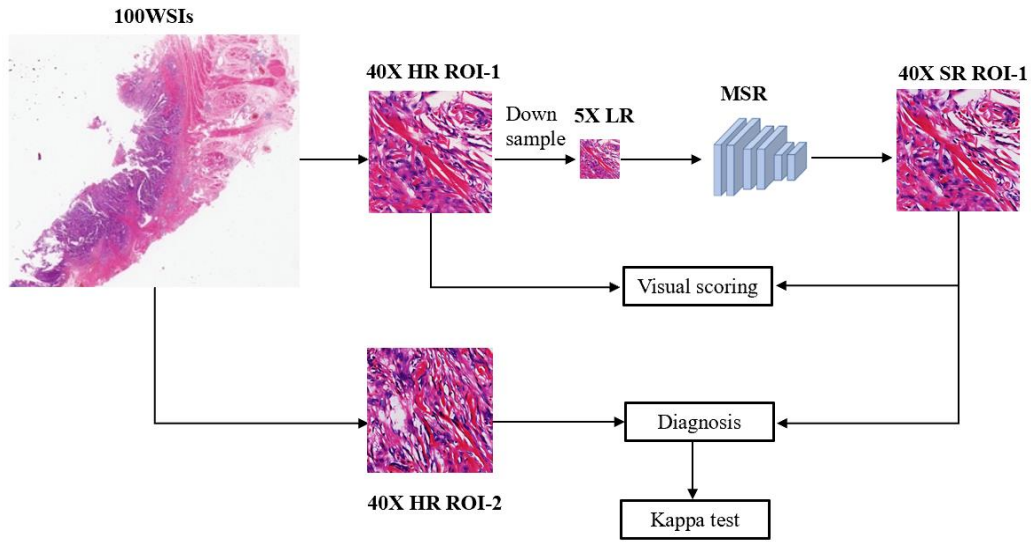

Figure 8. The flowchart of medical evaluation. Two ROIs from every WSI are selected, while ROI-1 is down sampled and then generated by our MSR method. The comparison of SR ROI-1 and HR ROI-1 by pathologist for visual scoring, and HR ROI-2 and SR ROI-1 are paired for consistent diagnosis experiment.

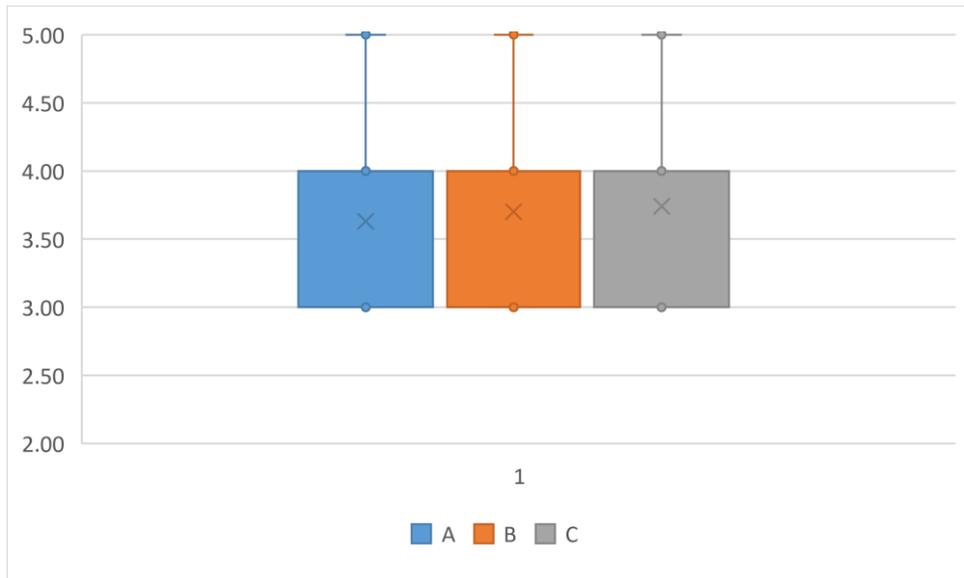

Figure 9. The box plot of visual scoring by pathologists A-C. The horizontal bar in the box indicates the median, while the cross indicates the mean.

| E | | HR ROI-2 | | | | | | | | | | | | | | | | | | |
|---|---|---|---|---|---|---|---|---|---|---|---|---|---|---|---|---|---|---|---|---|
| | | 0_C | 0_W | 1_C | 1_W | 2_C | 2_W | 3_C | 3_W | 4_C | 4_W | 5_C | 5_W | 6_C | 6_W | 7_C | 7_W | 8_C | 8_W | 9_C | 9_W |
| SR ROI-1 | 0_C | 10 | 0 | | | | | | | | | | | | | | | | | | |
| | 0_W | 0 | 0 | | | | | | | | | | | | | | | | | | |
| | 1_C | | | 10 | 0 | | | | | | | | | | | | | | | | |
| | 1_W | | | 0 | 0 | | | | | | | | | | | | | | | | |
| | 2_C | | | | | 10 | 0 | | | | | | | | | | | | | | |
| | 2_W | | | | | 0 | 0 | | | | | | | | | | | | | | |
| | 3_C | | | | | | | 10 | 0 | | | | | | | | | | | | |
| | 3_W | | | | | | | 0 | 0 | | | | | | | | | | | | |
| | 4_C | | | | | | | | | 10 | 0 | | | | | | | | | | |
| | 4_W | | | | | | | | | 0 | 0 | | | | | | | | | | |
| | 5_C | | | | | | | | | | | 10 | 0 | | | | | | | | |
| | 5_W | | | | | | | | | | | 0 | 0 | | | | | | | | |
| | 6_C | | | | | | | | | | | | | 10 | 0 | | | | | | |
| | 6_W | | | | | | | | | | | | | 0 | 0 | | | | | | |
| | 7_C | | | | | | | | | | | | | | | 10 | 0 | | | | |
| | 7_W | | | | | | | | | | | | | | | 0 | 0 | | | | |
| | 8_C | | | | | | | | | | | | | | | | | 10 | 0 | | |
| | 8_W | | | | | | | | | | | | | | | | | 0 | 0 | | |
| | 9_C | | | | | | | | | | | | | | | | | | | 10 | 0 |
| | 9_W | | | | | | | | | | | | | | | | | | | 0 | 0 |

(a)The diagnosis consistency of pathologist D

| D | | HR ROI-2 | | | | | | | | | | | | | | | | | | | |
|---|---|---|---|---|---|---|---|---|---|---|---|---|---|---|---|---|---|---|---|---|
| | | 0_C | 0_W | 1_C | 1_W | 2_C | 2_W | 3_C | 3_W | 4_C | 4_W | 5_C | 5_W | 6_C | 6_W | 7_C | 7_W | 8_C | 8_W | 9_C | 9_W |
| SR ROI-1 | 0_C | 9 | 0 | | | | | | | | | | | | | | | | | | |
| | 0_W | 1 | 0 | | | | | | | | | | | | | | | | | | |
| | 1_C | | | 10 | 0 | | | | | | | | | | | | | | | | |
| | 1_W | | | 0 | 0 | | | | | | | | | | | | | | | | |
| | 2_C | | | | | 10 | 0 | | | | | | | | | | | | | | |
| | 2_W | | | | | 0 | 0 | | | | | | | | | | | | | | |
| | 3_C | | | | | | | 10 | 0 | | | | | | | | | | | | |
| | 3_W | | | | | | | 0 | 0 | | | | | | | | | | | | |
| | 4_C | | | | | | | | | 10 | 0 | | | | | | | | | | |
| | 4_W | | | | | | | | | 0 | 0 | | | | | | | | | | |
| | 5_C | | | | | | | | | | | 10 | 0 | | | | | | | | |
| | 5_W | | | | | | | | | | | 0 | 0 | | | | | | | | |
| | 6_C | | | | | | | | | | | | | 10 | 0 | | | | | | |
| | 6_W | | | | | | | | | | | | | 0 | 0 | | | | | | |
| | 7_C | | | | | | | | | | | | | | | 10 | 0 | | | | |
| | 7_W | | | | | | | | | | | | | | | 0 | 0 | | | | |
| | 8_C | | | | | | | | | | | | | | | | | 10 | 0 | | |
| | 8_W | | | | | | | | | | | | | | | | | 0 | 0 | | |
| | 9_C | | | | | | | | | | | | | | | | | | | 10 | 0 |
| | 9_W | | | | | | | | | | | | | | | | | | | 0 | 0 |

(b) The diagnosis consistency of pathologist E

| F | | HR ROI-2 | | | | | | | | | | | | | | | | | | |
|---|---|---|---|---|---|---|---|---|---|---|---|---|---|---|---|---|---|---|---|---|
| | | 0_C | 0_W | 1_C | 1_W | 2_C | 2_W | 3_C | 3_W | 4_C | 4_W | 5_C | 5_W | 6_C | 6_W | 7_C | 7_W | 8_C | 8_W | 9_C | 9_W |
| SR ROI-1 | 0_C | 10 | 0 | | | | | | | | | | | | | | | | | | |
| | 0_W | 0 | 0 | | | | | | | | | | | | | | | | | | |
| | 1_C | | | 10 | 0 | | | | | | | | | | | | | | | | |
| | 1_W | | | 0 | 0 | | | | | | | | | | | | | | | | |
| | 2_C | | | | | 10 | 0 | | | | | | | | | | | | | | |
| | 2_W | | | | | 0 | 0 | | | | | | | | | | | | | | |
| | 3_C | | | | | | | 9 | 1 | | | | | | | | | | | | |
| | 3_W | | | | | | | 0 | 0 | | | | | | | | | | | | |
| | 4_C | | | | | | | | | 10 | 0 | | | | | | | | | | |
| | 4_W | | | | | | | | | 0 | 0 | | | | | | | | | | |
| | 5_C | | | | | | | | | | | 10 | 0 | | | | | | | | |
| | 5_W | | | | | | | | | | | 0 | 0 | | | | | | | | |
| | 6_C | | | | | | | | | | | | | 9 | 0 | | | | | | |
| | 6_W | | | | | | | | | | | | | 1 | 0 | | | | | | |
| | 7_C | | | | | | | | | | | | | | | 10 | 0 | | | | |
| | 7_W | | | | | | | | | | | | | | | 0 | 0 | | | | |
| | 8_C | | | | | | | | | | | | | | | | | 10 | 0 | | |
| | 8_W | | | | | | | | | | | | | | | | | 0 | 0 | | |
| | 9_C | | | | | | | | | | | | | | | | | | | 10 | 0 |
| | 9_W | | | | | | | | | | | | | | | | | | | 0 | 0 |

(c) The diagnosis consistency of pathologist F

Figure 10. The diagnosis consistency of pathologists D-F. The n (n=0, 1, 2...9) represents the n human body system that refers to Appendix A. The C and W represents the correct and wrong diagnosis respectively. The number in the table represents the number of consistently correct, consistently wrong, inconsistent diagnosis on the n human body system.

Appendix A

## Supplementary table 1. Types of diseases in each system

| ID | Human body systems | Types of disease | Numbers | |
|---|---|---|---|---|
| 0 | Head and neck | 1.Pleomorphic adenoma of salivary gland | 3 | |
| | | 2.Parotid gland cellular schwannoma | 3 | |
| | | 3.Papillary thyroid carcinoma | 5 | 20 |
| | | 4.Tongue squamous cell carcinoma | 2 | |
| | | 5.Laryngeal squamous cell carcinoma | 2 | |
| | | 6. Nodular thyroid cancer | 5 | |
| 1 | Intracranial nervous system | 1.Anaplastic oligodendroglioma | 2 | |
| | | 2.Anaplastic astrocytoma | 3 | |
| | | 3.Intracranial hair cell astrocytoma | 3 | 20 |
| | | 4.Intracranial meningioma | 7 | |
| | | 5.Hippocampal sclerosis | 5 | |
| 2 | Respiratory system | 1.Large cell neuroendocrine carcinoma of the lung | 6 | |
| | | 2.Lung adenocarcinoma | 2 | |
| | | 3.Lymph node metastatic lung adenocarcinoma | 2 3 | 20 |
| | | 4.Lung adenocarcinoma in situ | 4 | |
| | | 5.Keratinizing squamous cell carcinoma | 1 | |
| | | 6.Anterior mediastinal thymoma | 2 | |
| 3 | Digestive system | 1.Omentum Malignant Mesothelioma | 4 | |
| | | 2.Peritoneal mucous adenocarcinoma | 2 | |
| | | 3.Small intestine stromal tumor | 3 | |
| | | 4.Sigmoid colon adenocarcinoma | 2 | 20 |
| | | 5.Colon adenocarcinoma | 4 | |
| | | 6.Rectal adenocarcinoma | 5 | |
| 4 | Skin and soft tissue | 1.Poorly differentiated adenocarcinoma of eyelid | 3 | |
| | | 2.Highly differentiated myxoid liposarcoma of the buttocks | 6 4 | 20 |
| | | 3.Foot malignant melanoma | 2 | |
| | | 4.Embryonic rhabdomyosarcoma | 5 | |

Supplementary table 1. Types of diseases in each system(continued)

| ID | Human body systems | Types of disease | Numbers | |
|---|---|---|---|---|
| 5 | Female reproduction system | 1.Uterine Leiomyoma | 1 | 20 |
| | | 2.High-grade serous carcinoma of the uterus | 4 | |
| | | 3.Cervical endometrioid adenocarcinoma | 2 | |
| | | 4.Cervical adenocarcinoma in situ | 2 | |
| | | 5.Placenta adhesion and implantation | 3 | |
| | | 6.Uterine pregnancy | 1 | |
| | | 7.Cervical polyps | 2 | |
| | | 8.Endometrial polyps | 2 | |
| | | 9.Pelvic papillary serous carcinoma | 3 | |
| 6 | Urinary system | 1.Adrenal sebaceous adenoma | 4 | 20 |
| | | 2.Retroperitoneal leiomyosarcoma | 2 | |
| | | 3.Renal clear cell carcinoma | 2 | |
| | | 4.Reninoma | 3 | |
| | | 5.Kidney atrophy | 4 | |
| | | 6.Renal medullary lipoma | 5 | |
| 7 | Lymphoid hematopoietic system | 1.Reactive hyperplasia of peri-intestinal lymph nodes | 5 | 20 |
| | | 2.Reactive hyperplasia of inguinal lymph nodes | 4 | |
| | | 3.Reactive hyperplasia of perigastric lymph nodes | 5 | |
| | | 4.Ocular-MALT lymphoma | 6 | |
| 8 | Liver | 1.Hepatocellular carcinoma | 9 | 20 |
| | | 2.Nodular hyperplasia of liver | 11 | |
| 9 | Breast | 1.Invasive breast cancer | 20 | 20 |

Appendix B

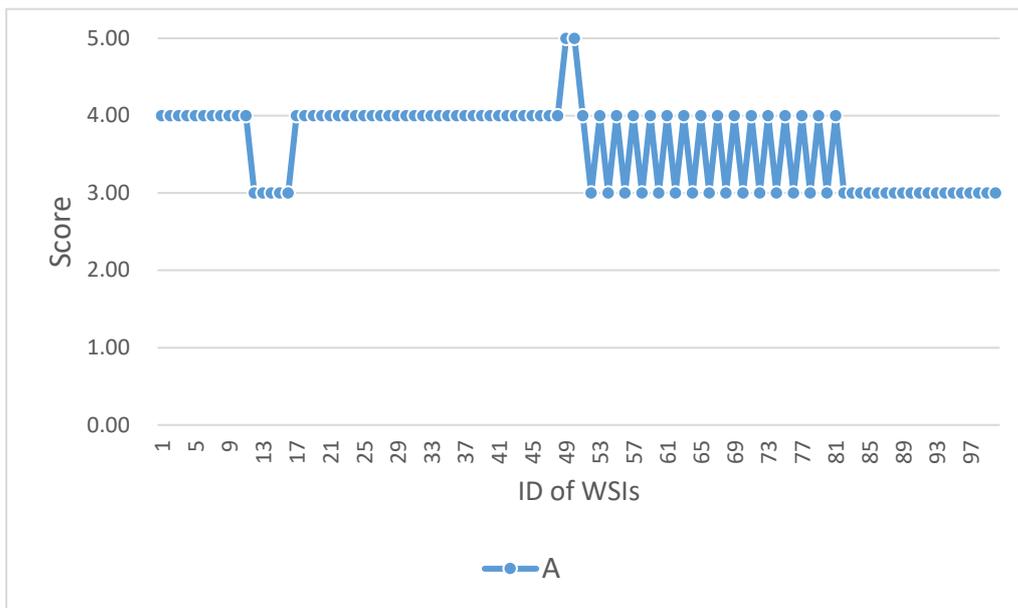

(a) The visual scoring of pathologists A

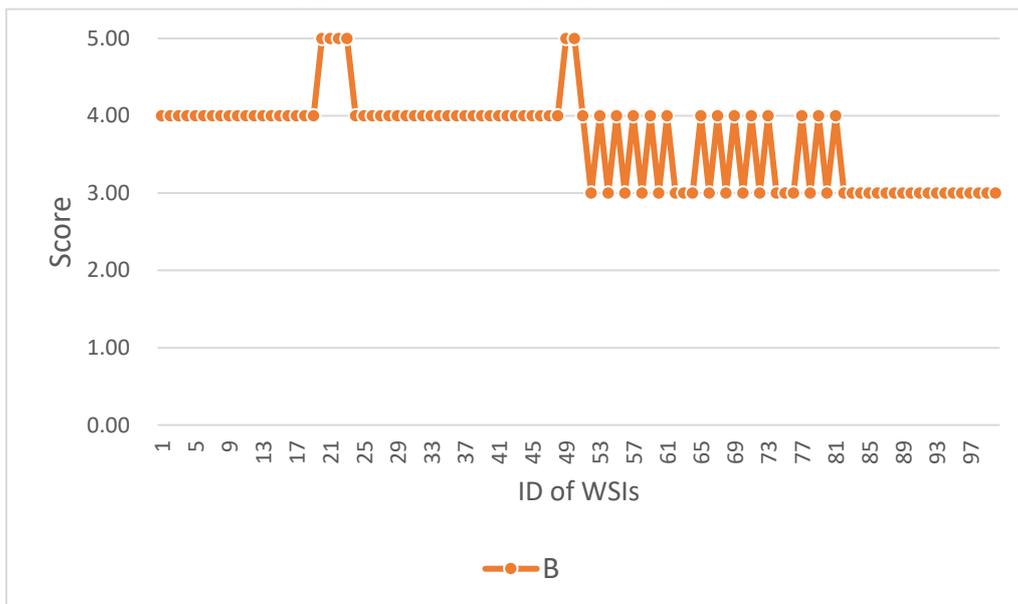

(b) The visual scoring of pathologists B

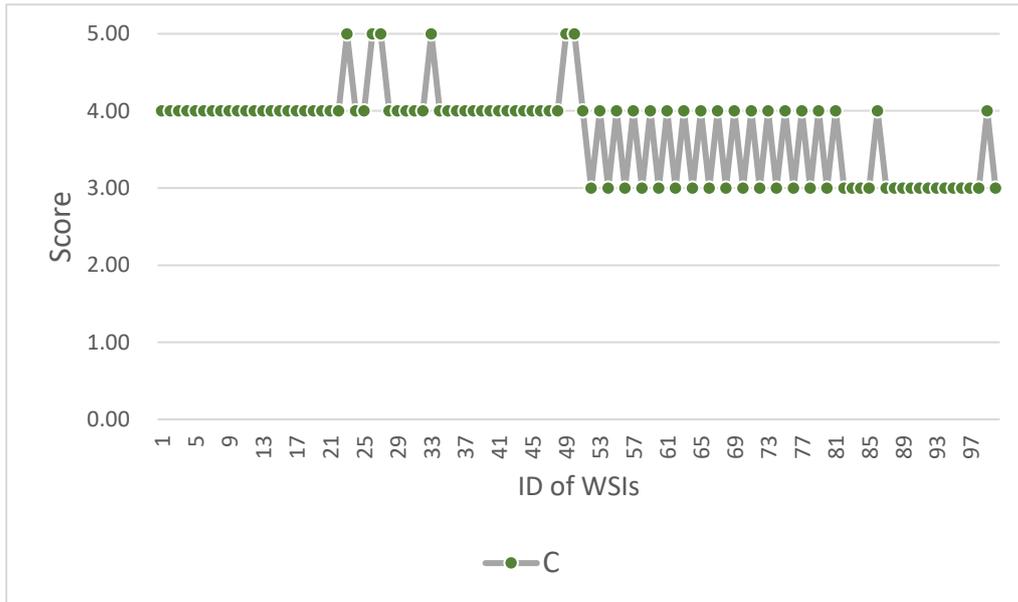

(c)The visual scoring of pathologists C

Supplementary figure 1. The three pictures (a), (b) and (c) respectively represent the visual scoring of the three pathologists A-C